\documentclass[twocolumn,aps,prl,preprintnumbers,superscriptaddress]{revtex4-2}
\usepackage{amsmath}
\usepackage{amssymb}
\usepackage{graphicx}
\usepackage{textcomp}
\usepackage{bm}
\usepackage[right]{eurosym}
\usepackage{float}
\usepackage[english]{babel}
\usepackage{blindtext}
\usepackage{booktabs}
\usepackage{longtable}
\usepackage{xcolor}
\usepackage[utf8]{inputenc}   
\usepackage[T1]{fontenc}      

\begin{document}

\author{N. Scheel}
\affiliation{Department of Physics and Astronomy, Aarhus University, 8000 Aarhus C, Denmark}

\author{N. S. Blaj}
\affiliation{Institute of Physics, University of Kassel, 34132 Kassel, Germany}

\author{R. Rajni}
\affiliation{Institute of Physics, University of Kassel, 34132 Kassel, Germany}

\author{A. Ø. Lægdsmand}
\affiliation{Institute of Physics, University of Kassel, 34132 Kassel, Germany}

\author{J. Sadasivan}
\affiliation{Department of Physics, Mahindra University Hyderabad, India}

\author{S. De}
\affiliation{Sorbonne Université, CNRS, Laboratoire de Chimie Physique Matière et Rayonnement, UMR 7614, F-75005 Paris, France}

\author{S. R. Krishnan}
\affiliation{Department of Physics and QuCenDiEM-group, Indian Institute of Technology Madras, Chennai 600036, India}

\author{J. Bozek}
\affiliation{Synchrotron SOLEIL, St. Aubin, BP48, 91192 Gif sur Yvette Cedex, France}

\author{A. R. Milosavljevi\'c}
\affiliation{Synchrotron SOLEIL, St. Aubin, BP48, 91192 Gif sur Yvette Cedex, France}

\author{M. Mudrich}\email{mudrich@uni-kassel.de}
\affiliation{Institute of Physics and CINSaT, University of Kassel, 34132 Kassel, Germany}

\title{Multiple photoelectron inelastic scattering in large He nanodroplets}

\begin{abstract}
Helium nanodroplets are ideal testbeds for elucidating secondary processes induced by ionizing radiation in condensed matter. Here we study photoionization electron spectra of large He nanodroplets in the extreme ultraviolet region, where photoelectrons undergo multiple inelastic collisions with He atoms in the nanodroplets. The resulting characteristic electron energy-loss features are identified by their dependence on photon energy and nanodroplet size and by comparison with a simple numerical simulation. Conditions for minimizing electron scattering with the He nanodroplets are discussed.
\end{abstract}
\date{\today}
\maketitle

\section{Introduction}
Inelastic scattering of electrons in condensed-phase environments plays a critical role in understanding fundamental processes of radiation-matter interaction such as energy dissipation, ionization, electron solvation, and radical formation. Helium nanodroplets have emerged as a valuable model system for investigating such phenomena due to their unique combination of properties. Comprised of weakly bound He atoms, these droplets exhibit a homogeneous, superfluid density distribution and offer an exceptionally clean and inert spectroscopic environment~\cite{Toennies2004}. The simplicity of the electronic structure of He and its electrophobicity, i.e. \, its tendency to repel electrons, enable the sensitive detection of high-resolution, easily interpretable electron spectra. This makes He nanodroplets an ideal platform for studying electron–matter interactions with minimal complication from internal electronic degrees of freedom~\cite{Shcherbinin2018,ltaief2023efficient,laforge2024interatomic}. Thus, He nanodroplets are well suited for unraveling electron scattering-induced secondary processes such as interatomic Coulombic decay (ICD)~\cite{ltaief2023efficient,Ltaief_2025,laforge2024interatomic}. These processes are fundamentally relevant for naturally occurring systems such as liquid and solid water in space, in the atmosphere, and in biology~\cite{boyer2016role,schmidt2022mechanisms, campbell2016electron, alizadeh2015biomolecular}. In particular, low-energy electrons have been found to efficiently cause radiation damage in living tissue by dissociative attachment to vital biomolecules~\cite{boudaiffa_resonant_2000,alizadeh2015biomolecular}. Therefore, systematic measurements using model systems such as He nanodroplets add valuable insight into secondary processes after photoionization in condensed matter systems.

This work focuses on inelastic scattering of energetic photoelectrons in large He nanodroplets that manifests itself as characteristic features in electron spectra at low energies. Using small He nanodroplets, we have previously mapped the signatures of inelastic photoelectron-He scattering~\cite{Shcherbinin2018}; photoelectron-impact excitation and ionization of He in the droplets gave rise to distinct peaks and broad components in the electron spectra at energies ranging between $E_e=0$ and 13~eV when tuning the photon energy between $h\nu=44$ and 56~eV. In large He nanodroplets, inelastic scattering of photoelectrons and subsequent trapping and recombination of the scattered electron with the parent photoion gave rise to metastable He atoms which subsequently decayed by ICD~\cite{ltaief2023efficient,Ltaief_2025}. At photon energies $h\nu > 65.4$~eV, He atoms are to some extent simultaneously ionized and excited, leading to ICD and the emission of a characteristic electron with energy $E_e\approx 8$~eV. 

Here, we extend our previous work by tuning the photon energy further up to a regime in which photoelectrons undergo multiple inelastic scattering. The resulting electron spectra turn out to feature additional characteristic structures which we analyze in detail. These results will aid in the interpretation of forthcoming x-ray photoelectron spectroscopic studies on pure and doped He nanodroplets using synchrotron radiation and x-ray free electron lasers.

\begin{figure*}[ht!]
    \centering
    \includegraphics[width=\textwidth]{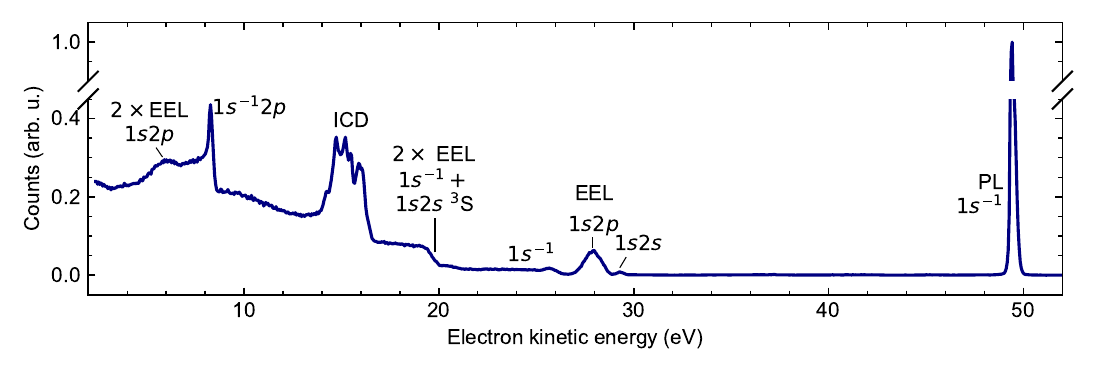}
    \caption{Photoelectron spectrum of large He nanodroplets produced at a nozzle temperature of 8~K ($R=75$~nm) recorded at a photon energy of $h\nu = 74$~eV. PL stands for photoline, EEL for electron energy loss, and ICD for interatomic Coulombic decay. The sharp peak at 8~eV labeled 1s$^{-1}$2p is due to shake-up ionization of He thus creating an excited He$^{+*}$ ion.}
    \label{fig:pes}
\end{figure*}

\begin{figure}[ht!]
    \centering
    \includegraphics[width=\linewidth]{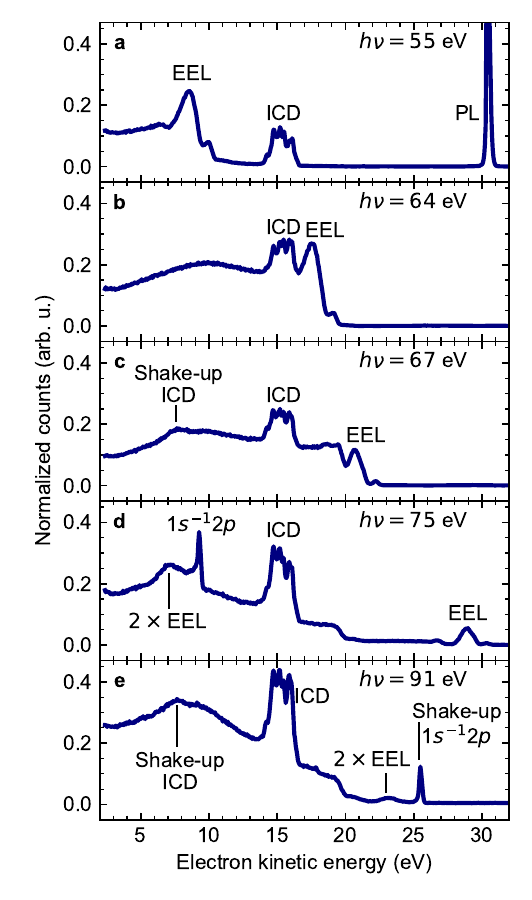}
    \caption{Photoelectron spectra of large He nanodroplets produced at a nozzle temperature of 8~K ($R=75$~nm) recorded at selected photon energies.
    }
    \label{fig:pes-55-75-eV}
\end{figure}

\section{Experimental Setup}
The experiments were carried out using a He nanodroplet source that has been described in detail elsewhere~\cite{Buchta2013,Shcherbinin2018,laforge2024interatomic}. Briefly, a continuous beam of He nanodroplets is generated by expanding He out of a cryogenic nozzle at a temperature ranging from 16~K to 8~K at a backing pressure of 50~bar. Under these conditions, He nanodroplets have variable sizes ranging from radii $R = 5$~nm (average number of He atoms per droplet $\langle N\rangle\sim 10^4$) to $R = 75$~nm ($\langle N\rangle\sim 10^8$), respectively. A transition from sub-critical expansion where small He nanodroplets ($R<6$~nm) are formed by condensation of gaseous He to super-critical expansion where large He nanodroplets ($R>6$~nm) are formed by fragmentation of liquid He behind the nozzle occurs when cooling the nozzle across the critical point ($T\approx 13$~K). Small He droplets produced in the sub-critical regime feature sharp lines in the electron spectra due to direct photoemission~\cite{peterka2007photoionization,Buchta2013a} and broader features due to inelastic electron-He scattering~\cite{shcherbinin2019inelastic}. The electron spectra of large He droplets formed by super-critical expansion contain additional structures induced by multiple elastic scattering, inelastic scattering, and electron-ion recombination giving rise to ICD~\cite{asmussen2023electron,ltaief2023efficient,ltaief2024spectroscopically,Ltaief_2025}. Additionally, the electron spectra of large He droplets may be distorted to some extent due to the tendency of large droplets to trap low-energy electrons in bubbles in the droplets resulting in missing signal in electron spectra~\cite{asmussen2023dopant}.

The He droplet source was attached to the PLEIADES beamline at the SOLEIL synchrotron facility~\cite{lindblad2013multi,danilovic2020aerosol}. A hemispherical electron analyzer
(HEA, model VG Scienta R4000) was used to record electron kinetic energy spectra. The HEA was equipped with a wide-angle lens mounted perpendicular to the photon propagation direction. To avoid distortions of the measured spectra due to angular anisotropies in the electron emission perpendicular to the photon propagation axis, circularly polarized radiation was used. The photon energy was tuned between $h\nu = 54$ and 110~eV, i.\,e. in a range where photoelectrons undergo one or multiple successive inelastic collisions with He atoms in the droplets. In most of the measurements presented in this study, a pass energy of 10~eV was used at the HEA. The exit slit of the monochromator in the beamline was set to 550~$\mu$m and the entrance slit of the HEA was 900~$\mu$m, resulting in an electron kinetic energy resolution of around 0.1~eV.

\begin{figure*}
    \centering
    \includegraphics{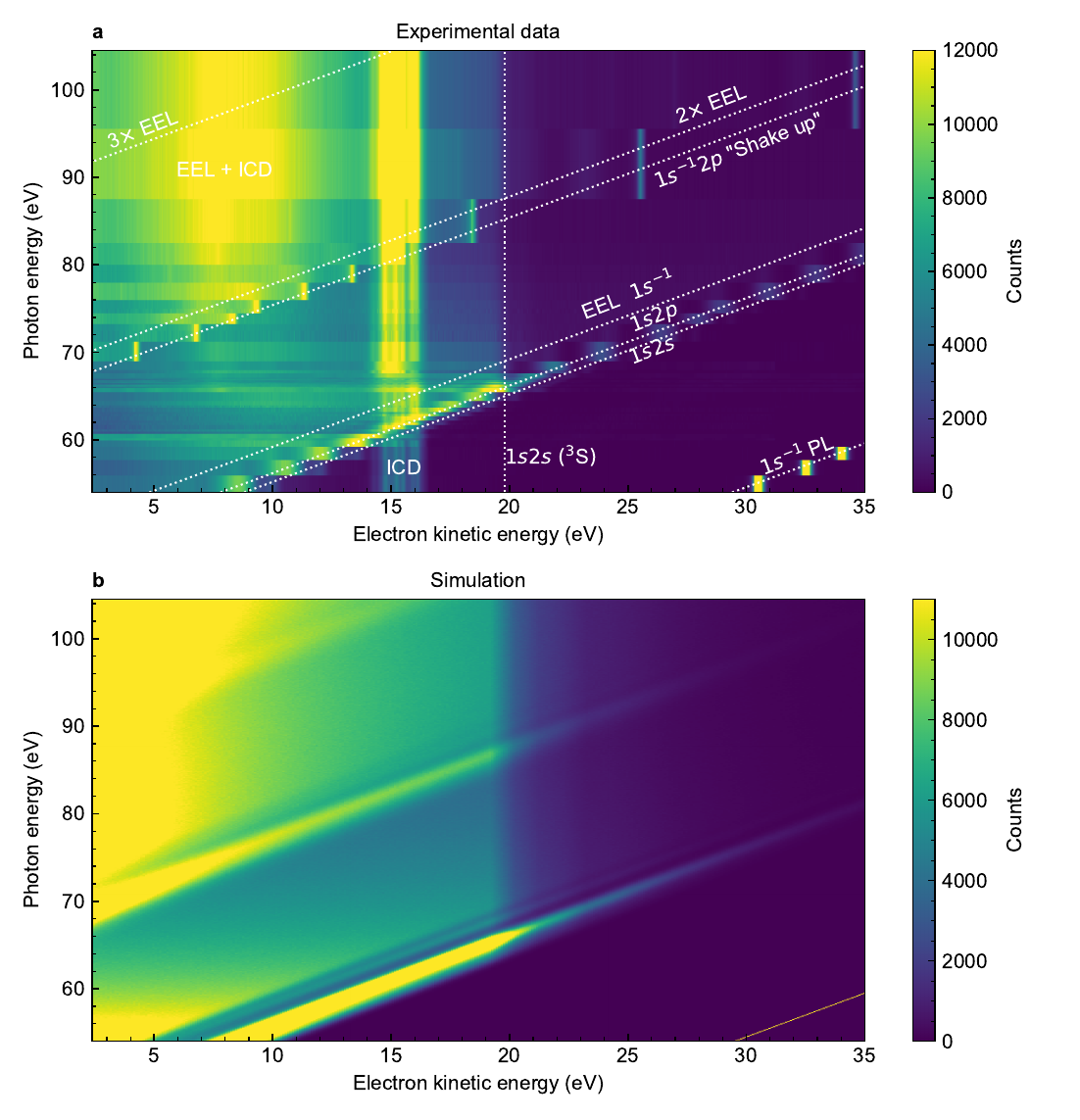}
    \caption{a) Density map showing measured electron spectra recorded for various photon energies. Dashed lines indicate characteristic energies calculated from literature values of the He atomic level energies. b) Simulated electron spectra. See text for details about the simulation.}
    \label{fig:2dmap}
\end{figure*}

\section{Results and Discussion}
A typical electron spectrum of large He nanodroplets ($R=75$~nm) irradiated at $h\nu = 74$~eV is shown in Fig.~\ref{fig:pes}. This spectrum gives a nice overview of the phenomenology of primary and secondary ionization processes that occur in He nanodroplets. The photoline due to photoionization of He (sharp peak at $E_e=h\nu-E_i^\mathrm{He} = 49.4$~eV) is accompanied by a satellite at $E_e=8.6$~eV due to the simultaneous ionization and excitation of a fraction of He atoms in a so-called shake-up process~\cite{svensson1988electron}. This process was previously found to induce ICD in He$_2$ dimers and in He nanodroplets giving rise to an additional distribution of ICD electrons in the range $E_e=5.5$-$19.5$~eV~\cite{Havermeier2010,Shcherbinin2017,laforge2024interatomic}. In the present case of large droplets, this ICD feature is completely covered by a broad peak structure resulting from inelastic scattering of photoelectron causing a loss of electron kinetic energy of $\gtrsim 20$~eV. The multi-peak structure around $E_e=15$~eV, which has previously been studied in detail~\cite{ltaief2023efficient,Ltaief_2025}, arises from ICD of metastable He atoms produced by electron impact excitation and electron-ion recombination in large He droplets.

The broad peak structure around $E_e=28$~eV accompanied by a continuous nearly constant band reaching down to $E_e=20$~eV is the electron-energy loss (EEL) spectrum of He nanodroplets; it is due to inelastic scattering of photoelectrons with He atoms in the droplet. The prominent peak at $E_e=27.8$~eV is attributed to impact excitation of the 1s2p\,$^1$P state in He droplets that has an excitation energy of 21.6~eV; the smaller peak at $E_e=29.2$~eV is due to impact excitation of the 1s2s\,$^3$S state in droplets (excitation energy about 20.2~eV), and the broad band at $20~\mathrm{eV}<E_e<25.6$~eV reflects electrons generated by impact ionization. The EEL spectra of small He nanodroplets have previously been discussed in detail~\cite{shcherbinin2019inelastic}. 

The new features we focus on in this work are the step at $E_e\approx 20$~eV and the broad structure at electron energies below this edge, which dominates the low-energy part of the electron spectra of large He droplets. These features are due to two successive inelastic collisions of photoelectrons in the droplets. The edge at $E_e\approx 20$~eV marks the cut-off above which primary electrons as well as secondary electrons, created by a first electron-He impact ionization event, can undergo another inelastic collision, thus exciting or ionizing another He atom in the same droplet and losing energy. This cut-off energy is given by the lowest possible He excitation, i.\,e. the 1s2s\,$^3$S state, that has an energy of 19.8~eV in the He atom. The maximum at $E_e=6.2$~eV is reminiscent of the peak at $E_e=27.8$~eV; it is due to a photoelectron that successively impact excites two He atoms in the 1s2p\,$^1$P state. 

To explore the photon energy dependence of these EEL and secondary ionization features, we recorded electron spectra at several photon energies in the range $54~\mathrm{eV}<h\nu<105$~eV, see the collection of electron spectra in Fig.~\ref{fig:pes-55-75-eV} and the density map in Fig.~\ref{fig:2dmap}. At photon energies $h\nu < 64.2$~eV, i.\,e. below the threshold for two successive electron-He inelastic collisions, the spectra contain only the photoline (PL), the characteristic EEL spectrum due to single electron-He collisions, and the ICD feature of impact-excited He atoms around $E_e=15$~eV. When the photon energy is scanned across this threshold, the amplitude of the main EEL peak due to single-collision 1s2p\,$^1$P excitation drops because part of the scattered electrons now undergo a second inelastic collision and further lose energy. As a result, two new features appear in the spectra -- a second peak $2\times 21.6$~eV$=43.2$~eV below the photoline and the edge at $E_e\approx 20$~eV. Above the threshold of shake-up ionization, the broad shake-up induced ICD feature appears in the same low kinetic energy region as the (multiple) EEL signal. 

The density map shown in Fig.~\ref{fig:2dmap} contains all measured spectra and, for reference, auxiliary dotted lines that indicate characteristic energies of directly emitted photoelectrons (PL), 1s$^{-1}2$p shake-up satellite, and EEL features due to 1, 2 and 3 successive inelastic electron-He collisions. The vertical dotted line at $E_e=19.8$~eV depicts the cut-off energy for inelastic collisions given by the lowest 1s2s\,$^3$S excitation of He atoms. As expected, the structure around $E_e=15$~eV, due to ICD which is an autoionization process, is independent of photon energy up to changes in amplitude. Most prominently, the intense signal at $E_e\leq 20$~eV, due to EEL by single and multiple successive electron-He collisions and to ICD after shake-up ionization (bright yellow area peaked around $E_e=8$~eV), is present in the entire range of photon energies $h\nu\gtrsim70$~eV. 

To support our interpretation of this prominent EEL feature, we have performed a simple one-dimensional classical Monte-Carlo simulation based on tabulated electron-He inelastic scattering cross sections~\cite{ralchenko2008electron}. In this simulation, we consider a He nanodroplet of radius $R=75$~nm with a uniform He density of $n=0.022$ atoms/Å$^3$. A primary electron is emitted in the center of the nanodroplet. This electron then propagates a distance $r$ until it inelastically scatters at a He atom or leaves the droplet without scattering, according to the probability density $\exp(-r/\ell)/\ell$ for a Poisson process in space. Here, $\ell=1/(n\sigma_\text{tot}(E_e))=12.4$~nm at the relevant electron kinetic energy $E_e=49.4$~eV is the mean free path of the electron inside the He droplet. The following inelastic scattering channels, which have the largest cross sections at $E_e = h\nu - E_i^\text{He}= 49.4$~eV, are taken into account: Impact excitation of He atomic states 1s2s$\,^{1,\,3}$S, 1s2p$\,^{1,\,3}$P, 1s3s$\,^{1,\,3}$S, 1s3p$\,^{1,\,3}$P, as well as impact ionization. In the latter case, a second electron is produced in the droplet center and propagated until the end of the simulation. The sharing of kinetic energy between the primary and secondary electrons is assumed to be uniform, which is a good approximation near threshold~\cite{Cvejanovic_1974}. 

As the energy of the primary electron, $E_e = 49.4$~eV, is high enough to undergo two successive inelastic collisions, after a first collision has taken place, the same procedure is repeated to allow a second inelastic collision to occur. Once the electron has left the volume of the He nanodroplet, the propagation is stopped, and the following information is stored in memory: Number of inelastic collisions (one or two), impact excitation or ionization, which He state(s) were excited in the case of impact excitation, and what kinetic energy the primary and secondary electrons have acquired in the case of impact ionization. This procedure is repeated for $10^6$ primary electrons. 

\begin{figure}
    \centering
    \includegraphics[width=1.0\columnwidth]{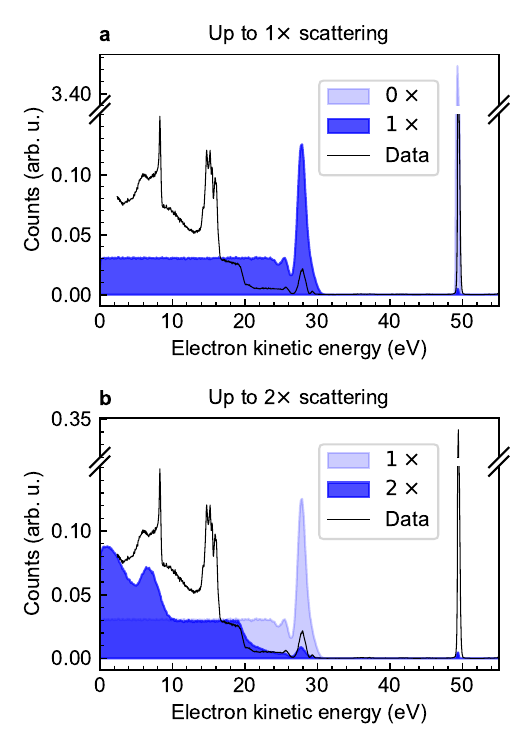}
    \caption{Experimental electron spectrum from Fig.~\ref{fig:pes} (black line) and simulated spectra for a photon energy $h\nu=74$~eV and droplet radius $R=75$~nm (blue areas). The histograms in a) show the simulated photoelectrons that have left the droplet without any scattering (light blue area near $E_e=50$~eV) and those that have undergone one inelastic collision (dark blue area). Panel b) compares the single-collision spectrum from a) (light blue area) to the simulated spectrum that includes a second collision (dark blue). The edge feature at $E_e=20$~eV and the low-energy part of the experimental spectrum at $E_e<10$~eV (except the shake-up satellite peak at $E_e<9$~eV) are well reproduced by the simulation including up to two inelastic electron-He collisions. 
    }
    \label{fig:2collisions}
\end{figure}

The result is represented as a histogram as a function of electron energy. The spectrum of EEL peaks is convolved with a Gaussian function with a $\mathrm{FWHM}= 1.2$~eV to account for the intrinsic broadening of the electronic energy levels in He nanodroplets~\cite{shcherbinin2019inelastic}, see the shaded areas in Fig.~\ref{fig:2collisions}. A more detailed discussion of the simulation and the resulting histograms for individual scattering channels is given in the Supplementary Material (SM).
The simulation that takes into account only one collision, shown in panel a), reproduces the peak at $E_e\approx 28$~eV in the experimental spectrum (1s2p excitation) and a continuous distribution at $E_e<26$~eV (electron-impact ionization). Clearly, this simulated distribution does not match the experimental spectrum at $E_e\lesssim 20$~eV very well. However, when a second successive inelastic collision is included (panel b), the main structure of the secondary electron spectrum ($E_e<30$~eV) is captured surprisingly well. Naturally, since shake-up ionization and ICD are not accounted for in the simulation, the corresponding features around 8 and 15~eV, respectively, are not reproduced. When these features are ignored, the experimental and the simulated spectra agree quite well. Especially the edge around $E_e\approx 20$~eV is reproduced, supporting our interpretation that electrons with a kinetic energy higher than the lowest He excitation energy will likely undergo an inelastic collision and shift down in energy. Thus, inelastic electron-He scattering plays an important role in He nanodroplet experiments whenever electrons are produced with kinetic energy $E_e\gtrsim 20$~eV.

By repeating this simulation for various values of the photon energy, we obtain a two-dimensional map of simulated electron spectra to be compared with the experimental data, see Fig.~\ref{fig:2dmap}~b). Overall, it reproduces the main features indicative of inelastic scattering -- the EEL peak due to single excitation of mainly the 1s2p\,$^1$P state of He nanodroplets (second lowest dashed slanted line), the cut-off at $E_e\approx 20$~eV above which the signal level generally drops (dashed vertical line), and the broad contribution of low-energy electrons at photon energies $>70$~eV due to two or more successive inelastic collisions (``$2\times$EEL'', ``$3\times$EEL''). Note that in that region, the latter overlaps with ICD electrons created with kinetic energies $5.5\lesssim E_e \lesssim 19.5$~eV by shake-up ionization. Because the latter are not included in the simulation, the agreement between the experiment and the simulation remains imperfect in this region.

\begin{figure}
    \centering
    \includegraphics{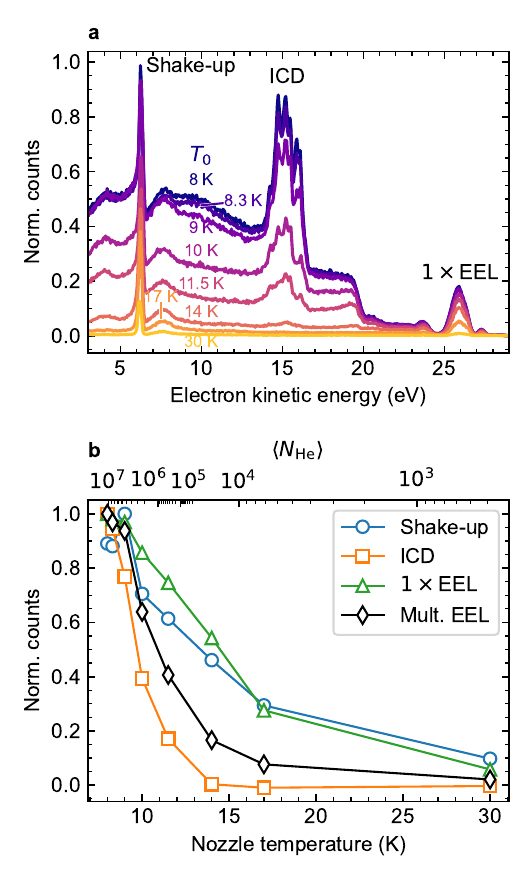}
    \caption{a) Electron spectra recorded at $h\nu =72$~eV for various nozzle temperatures from 30~K down to 8~K. b) Peak integrals of the various spectral features in a) as a function of nozzle temperature which controls the He nanodroplet size.}
    \label{fig:temps}
\end{figure}
To systematically characterize the dependence of the inelastic scattering features on the He droplet size, we recorded a series of electron spectra at $h\nu = 72$~eV for variable He nozzle temperature between $T_0=30$~K and 8~K. The corresponding droplet sizes are given as tick labels on the top horizontal axis of Fig.~\ref{fig:temps}~b). For small He nanodroplets with $\langle N\rangle\lesssim 4\times 10^4$ ($T_0\geq 14$~K), the EEL peak due to a single electron-He inelastic collision at $E_e=26$~eV and the shake-up satellite at $E_e=6$~eV are the dominant features. However, as the droplets grow much larger at $T_0\leq 11.5$~K, the additional broad spectral structure at $E_e<20$~eV and the ICD feature around $E_e=15$~eV become more prominent, while the EEL peak at $E_e=26$~eV increases more slowly. 

The droplet size dependence of the different spectral components is quantitatively analyzed by fitting the spectra with the sum of Gaussian functions. See the SM for details of the fit procedure. The resulting peak areas normalized to their maximum values at low temperature are shown in Fig.~\ref{fig:temps}~b). The single-collision peak (`$1\times$EEL', green triangles) decreases from its maximum value at $T_0<10~$K to about one half at $T_0\approx 15~$K and to nearly zero at $T_0=30~$K mainly due to the decreasing density of He atoms in the interaction region. The narrow satellite peak associated with shake-up ionization of He atoms (blue symbols) shows a similar decrease, although its value at $T_0=30~$K remains significantly larger, $\approx 0.1$. This peak can be taken as an approximate measure of the overall He density in the interaction region. Note, however, that the fit of this peak to determine its integral is subject to substantial uncertainties due to several overlapping spectral components that need to be subtracted, see the SM. In a previous study, the ratio of EEL peak integral to the photoline was found to systematically decrease and to vanish in the limit of very small He nanodroplets~\cite{shcherbinin2019inelastic}. We expect the same trend in the present experiment.

More importantly, the broad multiple-scattering structure (black symbols) and the ICD feature (orange) exhibit an even steeper drop when $T_0$ increases from 8 to 18~K. Note that the former contains contributions from both single collisions leading to impact ionization (flat distribution from $E_e=0$ to 26~eV, see Fig.~\ref{fig:2collisions}~a), two inelastic collisions and ICD following shake-up ionization~\cite{Shcherbinin2017}. As expected, that component varies with droplet size in a similar way as the ICD feature; Both features (multiple-scattering and ICD) rely on at least two inelastic collisions occurring in the same droplet before the electron escapes from the droplet~\cite{ltaief2023efficient}. 

Another set of electron spectra for varying He nozzle temperature was recorded at $h\nu = 58$~eV, see Fig.~S5 in the SM. Since at that energy, photoelectrons can only undergo one inelastic collision, the spectra are dominated by that respective EEL feature, aside from the photoline at $E_e=33.4$~eV; The prominent step at $E_e\approx 20$~eV due to two successive inelastic collisions, seen at $h\nu = 72$~eV, is absent. The integrals of the remaining peaks (photoline, ICD and single-collision EEL feature) follow a similar dependence on the nozzle temperature as that shown in Fig.~\ref{fig:temps}~b): While the photoline slowly drops in intensity as the nozzle temperature $T_0$ increases, the EEL signal (single collision) drops slightly faster and reaches nearly zero for $T_0$ ranging from 8 to 17~K. The ICD signal drops to nearly zero significantly faster in the range $T_0=8$ to 12~K.

\section{Conclusions and outlook}
In summary, photoelectron spectra of large He nanodroplets studied by high-resolution photoelectron spectroscopy in the XUV spectral range show pronounced structures due to multiple inelastic electron-He collisions. Overall, various superimposing features from secondary processes are well discernible, including shake-up ionization and ICD in addition to single and multiple electron-He inelastic scattering. Two successive electron-He collisions manifest themselves as a step-like feature at an electron energy of about 20~eV and a broad maximum peaked about 44~eV below the photoline. These prominent features are present in electron spectra up to a photon energy of 105~eV and are likely present at much higher photon energies as well. They will have to be taken into account in any future attempts to measure photoelectron spectra of molecules or complexes embedded in He nanodroplets using XUV and soft x-ray radiation. Such multiple inelastic scattering processes likely play an important role in other condensed-phase systems such as liquid water as well~\cite{signorell2022photoionization}. As they generate low-energy electrons, they can efficiently induce further secondary reactions causing radiation damage such as dissociation of biomolecules in biological matter~\cite{alizadeh2015biomolecular}.

Possible routes to minimizing the effects of electron-He inelastic scattering are the use of small He nanodroplets, as the broad spectral feature due to multiple scattering is nearly absent for droplets containing $\lesssim 10^4$ He atoms. However, in such nanodroplets only small dopant complexes composed of $\lesssim10$ atoms or molecules can be formed. Alternatively, x-rays with a much higher photon energy $\gtrsim 1$~keV could be used; the electron impact ionization cross section steeply drops for electron energies $>150$~eV~\cite{ralchenko2008electron}. However, this requires the use of hard x-rays and an electron analyzer capable of detecting correspondingly energetic electrons. Another approach could be to deliberately shrink the shell of He atoms surrounding the embedded species by heavily loading the droplets with the dopant species; the heat of cluster formation released into the nanodroplet can lead to nearly complete evaporation of all He leaving behind the bare dopant cluster~\cite{Blaj_2026}. Recently, controlled scattering of He nanodroplets doped with charged species in a separate He-filled scattering cell was shown to shrink or completely evaporate the He shell that hosts the dopant molecule or cluster~\cite{bergmeister2023spectroscopy}. Whether this method can be applied to He nanodroplets doped with neutral species remains to be demonstrated.

\section*{Supplementary Material}
The Supplementary Material provides additional details on the numerical simulation and analysis of electron energy-loss spectra presented in the main text. The energy-dependent electron-He inelastic scattering cross sections used in the Monte Carlo model are discussed, and the resulting probabilities for the individual excitation and impact-ionization channels are presented. The contributions of successive inelastic scattering events to the simulated electron spectra are analyzed in detail. We further describe the Poissonian treatment of scattering events in the nanodroplets and discuss the assumptions and limitations of the simplified simulation model. In addition, the fitting procedures used to extract the contributions of electron energy-loss processes, ICD, and shake-up features from the experimental spectra are presented. Finally, measurements in the single-collision regime are used to further characterize the dependence of the observed spectral features on the He nanodroplet size.

\section*{Acknowledgements}
We acknowledge SOLEIL for provision of synchrotron radiation facilities, and we thank the PLEIADES staff for assistance in preparing the experiment and using the beamline under proposal 20241684. Financial support by the Novo Nordisk Foundation (grant no. NNF23OC0085401) and the Deutsche Forschungsgemeinschaft (DFG, German Research Foundation) – SFB1319 – Projektnummer 328961117 is gratefully acknowledged. S.R.K. and J.S. thank for the support by the Indo-French Center for Promotion of Academic Research (CEFIPRA). The research leading to these results has been supported by the COST Action CA21101 ``Confined Molecular Systems: From a New Generation of Materials to the Stars (COSY)''. 

\bibliography{Bib}

@String{jcp={J. Chem. Phys.}}

@String{jpca={J. Phys. Chem. A}}

@String{jpcc={J. Phys. Chem. C}}

@String{rsi={Rev. Sci. Instrum.}}

@String{s={Science}}

@Article{Buchta2013,
  Author                   = {Buchta, Dominic and Krishnan, Siva R. and Brauer, Nils B. and Drabbels, Marcel and O'Keeffe, Patrick and Devetta, Michele and Di Fraia, Michele and Callegari, Carlo and Richter, Robert and Coreno, Marcello and others},
  Title                    = {Charge transfer and penning ionization of dopants in or on helium nanodroplets exposed to EUV radiation},
  Journal                  = {J. Phys. Chem. A},
  Year                     = {2013},
  Volume                   = {117},
  Number                   = {21},
  Pages                    = {4394--4403},
  Publisher                = {ACS Publications},
}

@Article{Buchta2013a,
  Author                   = {Buchta, D. and Krishnan, Siva Rama and Brauer, Nils Benedict and Drabbels, Marcel and O'Keeffe, P. and Devetta, M. and Di Fraia, Michele and Callegari, C. and Richter, R. and Coreno, M. and others},
  Title                    = {Extreme ultraviolet ionization of pure He nanodroplets: Mass-correlated photoelectron imaging, Penning ionization, and electron energy-loss spectra},
  Journal                  = {J. Chem. Phys.},
  Year                     = {2013},
  Volume                   = {139},
  Number                   = {8},
  Pages                    = {084301},
  Publisher                = {AIP},
}

@Article{peterka2007photoionization,
  Title                    = {Photoionization dynamics in pure helium droplets},
  Author                   = {Peterka, Darcy S. and Kim, Jeong Hyun and Wang, Chia C. and Poisson, Lionel and Neumark, Daniel M.},
  Journal                  = jpca,
  Volume                   = {111},
  Number                   = {31},
  Pages                    = {7449--7459},
  Year                     = {2007},
  Publisher                = {ACS Publications}
}

@Article{Toennies2004,
  Title                    = {Superfluid helium droplets: A uniquely cold nanomatrix for molecules and molecular complexes},
  Author                   = {Toennies, J. Peter and Vilesov, Andrey F.},
  Year                     = {2004},
  Number                   = {20},
  Pages                    = {2622},
  Volume                   = {43},
  Journal                  = {Angew. Chem. Int. Ed.},
  Publisher                = {Wiley Online Library},
  Doi                      = {10.1002/anie.200300611},
}

@Article{Shcherbinin2017,
  Author                   = {Shcherbinin, Mykola and LaForge, A. C. and Sharma, Vandana and Devetta, Michele and Richter, Robert and Moshammer, Robert and Pfeifer, Thomas and Mudrich, Marcel},
  Title                    = {Interatomic {C}oulombic decay in helium nanodroplets},
  Journal                  = {Phys. Rev. A},
  Year                     = {2017},
  Volume                   = {96},
  Number                   = {1},
  Pages                    = {013407}
}

@Article{Havermeier2010,
  Author                   = {Havermeier, T. and Jahnke, T. and Kreidi, K. and Wallauer, R. and Voss, S. and Sch\"{o}ffler, M. and Sch\"{o}ssler, S. and Foucar, L. and Neumann, N. and Titze, J. and others},
  Title                    = {Interatomic {C}oulombic decay following photoionization of the helium dimer: Observation of vibrational structure},
  Journal                  = {Phys. Rev. Lett.},
  Year                     = {2010},
  Volume                   = {104},
  Number                   = {13},
  Pages                    = {133401}
}

@article{ltaief2024spectroscopically,
  title={Spectroscopically resolved resonant interatomic Coulombic decay in photoexcited large He nanodroplets},
  author={Ben Ltaief, L. and Sishodia, K. and Richter, R. and Bastian, B. and Asmussen, J. D. and Mandal, S. and Pal, N. and Medina, C. and Krishnan, S. R. and von Haeften, K. and Mudrich, M.},
  journal={Phys. Rev. Research},
  volume={6},
  number={1},
  pages={013019},
  year={2024}
}

@article{Ltaief_2025,
year = {2025},
month = {feb},
volume = {88},
number = {3},
pages = {037901},
author = {Ben Ltaief, L and Sishodia, K and Asmussen, J D and Abid, A R and Krishnan, S R and Pedersen, H B and Sisourat, N and Mudrich, M},
title = {Interatomic Coulombic decay in lithium-doped large helium nanodroplets induced by photoelectron impact excitation},
journal = {Rep. Prog. Phys.}
}

@article{laforge2024interatomic,
  title={Interatomic and intermolecular decay processes in quantum fluid clusters},
  author={LaForge, Aaron Christopher and Ltaief, L Ben and Krishnan, SR and Sisourat, Nicolas and Mudrich, Marcel},
  journal = {Rep. Prog. Phys.},
  volume={87},
  number={12},
  pages={126402},
  year={2024},
  publisher={IOP Publishing}
}

@Article{Shcherbinin2018,
  Author                   = {Shcherbinin, M. and LaForge, AC and Hanif, M. and Richter, R. and Mudrich, M.},
  Title                    = {Penning ionization of acene molecules by helium nanodroplets},
  Journal                  = {J. Phys. Chem. A},
  Year                     = {2018},
  Volume                   = {122},
  Number                   = {7},
  Pages                    = {1855--1860}
}

@Article{boudaiffa_resonant_2000,
  Title                    = {Resonant {Formation} of {DNA} {Strand} {Breaks} by {Low}-{Energy} (3 to 20 {eV}) {Electrons}},
  Volume                   = {287},
  ISSN                     = {0036-8075, 1095-9203},
  Doi                      = {10.1126/science.287.5458.1658},
  Language                 = {en},
  Number                   = {5458},
  Journal                  = {Science},
  Author                   = {Boudaïffa, Badia and Cloutier, Pierre and Hunting, Darel and Huels, Michael A. and Sanche, L\'{e}on},
  Month                    = {mar},
  Year                     = {2000},
  Pmid                     = {10698742},
  Pages                    = {1658--1660}
}

@Article{shcherbinin2019inelastic,
  Title                    = {Inelastic scattering of photoelectrons from {He} nanodroplets},
  Author                   = {Shcherbinin, M. and Westergaard, F. Vad and Hanif, M. and Krishnan, SR and LaForge, AC and Richter, R. and Pfeifer, Thomas and Mudrich, M.},
  Journal                  = jcp,
  Volume                   = {150},
  Number                   = {4},
  Pages                    = {044304},
  Year                     = {2019},
  Publisher                = {AIP Publishing LLC},
  Doi                      = {10.1063/1.5074130},
}

@Article{toennies_superfluid_2004,
  Title                    = {Superfluid {Helium} {Droplets}: {A} {Uniquely} {Cold} {Nanomatrix} for {Molecules} and {Molecular} {Complexes}},
  Volume                   = {43},
  ISSN                     = {1433-7851, 1521-3773},
  Shorttitle               = {Superfluid {Helium} {Droplets}},
  Doi                      = {10.1002/anie.200300611},
  Language                 = {en},
  Number                   = {20},
  Urldate                  = {2018-07-03},
  Journal                  = {Angew. Chem. Int. Ed.},
  Author                   = {Toennies, J. Peter and Vilesov, Andrey F.},
  Month                    = {may},
  Year                     = {2004},
  Pages                    = {2622--2648}
}

@String{ft={J.~Chem.\ Soc., Faraday Trans.}}

@String{jcp={J.~Chem.\ Phys.}}

@String{jpca={J.~Phys.\ Chem.~A}}

@Article{ralchenko2008electron,
  Title                    = {Electron-impact excitation and ionization cross sections for ground state and excited helium atoms},
  Author                   = {Ralchenko, Yu and Janev, RK and Kato, T. and Fursa, DV and Bray, Igor and De Heer, FJ},
  Journal                  = {At. Data Nucl. Data Tables},
  Volume                   = {94},
  Number                   = {4},
  Pages                    = {603--622},
  Year                     = {2008},
  Publisher                = {Elsevier},
  Doi                      = {10.1016/j.adt.2007.11.003},
}

@Article{ltaief2023efficient,
  Author                   = {Ben Ltaief, L. and Sishodia, K. and Mandal, S. and De, S. and Krishnan, S. R. and Medina, C. and Pal, N. and Richter, R. and Fennel, T. and Mudrich, M.},
  Title                    = {Efficient Indirect Interatomic {C}oulombic Decay Induced by Photoelectron Impact Excitation in Large Pure Helium Nanodroplets},
  Journal                  = {Phys. Rev. Lett.},
  Year                     = {2023},
  Volume                   = {131},
  Pages                    = {023001},
  Month                    = {Jul},
  Doi                      = {10.1103/PhysRevLett.131.023001},
  Issue                    = {2},
  Numpages                 = {7},
  Publisher                = {American Physical Society}
}

@Article{asmussen2023dopant,
  Author                   = {Asmussen, Jakob D. and Ben Ltaief, Ltaief and Sishodia, Keshav and Abid, Abdul R. and Bastian, Bj\"{o}rn and Krishnan, Sivarama and Pedersen, Henrik B. and Mudrich, Marcel},
  Title                    = {Dopant ionization and efficiency of ion and electron ejection from helium nanodroplets},
  Journal                  = {J. Chem. Phys.},
  Year                     = {2023},
  Volume                   = {159},
  Number                   = {3},
  Pages                    = {034301},
  Month                    = {07},
  ISSN                     = {0021-9606},
  Doi                      = {10.1063/5.0160171}
}

@Article{asmussen2023electron,
  Author                   = {Asmussen, Jakob D. and Sishodia, Keshav and Bastian, Bj\"{o}rn and Abid, Abdul R. and Ben Ltaief, Ltaief and Pedersen, Henrik B. and De, Subhendu and Medina, Cristian and Pal, Nitish and Richter, Robert and Fennel, Thomas and Krishnan, Sivarama and Mudrich, Marcel},
  Title                    = {Electron energy loss and angular asymmetry induced by elastic scattering in superfluid helium nanodroplets},
  Journal                  = {Nanoscale},
  Year                     = {2023},
  Volume                   = {15},
  Pages                    = {14025--14031},
  Doi                      = {10.1039/D3NR03295G},
  Issue                    = {34},
  Publisher                = {The Royal Society of Chemistry}
}

@article{svensson1988electron,
  title={Electron shake-up and correlation satellites and continuum shake-off distributions in X-Ray photoelectron spectra of the rare gas atoms},
  author={Svensson, Svante and Eriksson, Bengt and M{\aa}rtensson, Nils and Wendin, G{\"o}ran and Gelius, Ulrik},
  journal={J. Electron Spectrosc. Relat. Phenom.},
  volume={47},
  pages={327--384},
  year={1988}
}

@article{lindblad2013multi,
  title={A multi purpose source chamber at the PLEIADES beamline at SOLEIL for spectroscopic studies of isolated species: Cold molecules, clusters, and nanoparticles},
  author={Lindblad, Andreas and S{\"o}derstr{\"o}m, Johan and Nicolas, Christophe and Robert, Emmanuel and Miron, Catalin},
  journal=rsi,
  volume={84},
  number={11},
  year={2013}
}

@article{danilovic2020aerosol,
  title={Aerosol synthesis and gas-phase photoelectron spectroscopy of ag-bi-i nanosystems},
  author={Danilovi\'{c}, Danijela and Bo\v{z}ani\'{c}, Du\v{s}an K and Doj\v{c}ilovi\'{c}, Radovan and Vukmirovi\'{c}, Nenad and Sapkota, Pitambar and Vuka\v{s}inovi\'{c}, Ivana and Djokovi\'{c}, Vladimir and Bozek, John and Nicolas, Christophe and Ptasinska, Sylwia and others},
  journal=jpcc,
  volume={124},
  number={43},
  pages={23930--23937},
  year={2020},
  publisher={ACS Publications}
}

@article{bergmeister2023spectroscopy,
  title={Spectroscopy of helium-tagged molecular ions—Development of a novel experimental setup},
  author={Bergmeister, Stefan and Ganner, Lisa and Locher, Johannes and Zappa, Fabio and Scheier, Paul and Gruber, Elisabeth},
  journal={Rev. Sci. Instrum.},
  volume={94},
  number={5},
  year={2023},
  publisher={AIP Publishing}
}

@article{signorell2022photoionization,
  title={Photoionization of the aqueous phase: clusters, droplets and liquid jets},
  author={Signorell, Ruth and Winter, Bernd},
  journal=pcccp,
  volume={24},
  number={22},
  pages={13438--13460},
  year={2022},
  publisher={Royal Society of Chemistry}
}

@article{alizadeh2015biomolecular,
  title={Biomolecular damage induced by ionizing radiation: the direct and indirect effects of low-energy electrons on DNA},
  author={Alizadeh, Elahe and Orlando, Thomas M and Sanche, L{\'e}on},
  journal={Annu. Rev. Phys. Chem.},
  volume={66},
  number={1},
  pages={379--398},
  year={2015},
  publisher={Annual Reviews}
}

@article{schmidt2022mechanisms,
  title={Mechanisms of electron-induced chemistry in molecular ices},
  author={Schmidt, Fabian and Borrmann, Tobias and Mues, Martin Philipp and Benter, Sanna and Swiderek, Petra and Bredeh{\"o}ft, Jan Hendrik},
  journal={Atoms},
  volume={10},
  number={1},
  pages={25},
  year={2022},
  publisher={MDPI}
}

@article{boyer2016role,
  title={The role of low-energy (≤ 20 eV) electrons in astrochemistry},
  author={Boyer, Michael C and Rivas, Nathalie and Tran, Audrey A and Verish, Clarissa A and Arumainayagam, Christopher R},
  journal={Surface Science},
  volume={652},
  pages={26--32},
  year={2016},
  publisher={Elsevier}
}

@article{campbell2016electron,
  title={Electron collisions in atmospheres},
  author={Campbell, Laurence and Brunger, MJ},
  journal={International Reviews in Physical Chemistry},
  volume={35},
  number={2},
  pages={297--351},
  year={2016},
  publisher={Taylor \& Francis}
}

@article{Cvejanovic_1974,
    doi = {10.1088/0022-3700/7/14/008},
    url = {https://doi.org/10.1088/0022-3700/7/14/008},
    year = {1974},
    month = {oct},
    publisher = {},
    volume = {7},
    number = {14},
    pages = {1841},
    author = {S Cvejanovic and F H Read},
    title = {Studies of the threshold electron impact ionization of helium},
    journal = {Journal of Physics B: Atomic and Molecular Physics}
}

@Article{Blaj_2026,
    Author                   = {N. S. Blaj and N. Scheel and R. Rajni and A. {\O}. L\ae{}gdsmand and J. Sadasivan and S. De and S. R. Krishnan and J. Bozek and A. R. Milosavljevi{\'c} and M. Mudrich},
    Year                     = {2026},
    Note                     = {arXiv: 2609.04553},
    Title                    = {X-ray photoelectron spectroscopy of Ar and Kr clusters formed in He nanodroplets},
    Url                      = {https://arxiv.org/pdf/2609.04553},
    Journal                  = {}
}

\end{document}


\begin{center}
{\bf Supplementary Information for:}\\[0.5em]
{\bf Multiple photoelectron inelastic scattering in large helium nanodroplets}\\[1em]
N. Scheel \textit{et al.}
\end{center}

\vspace{1em}

\section{Numerical simulation}
\begin{figure}[hb!]
    \centering
    \includegraphics{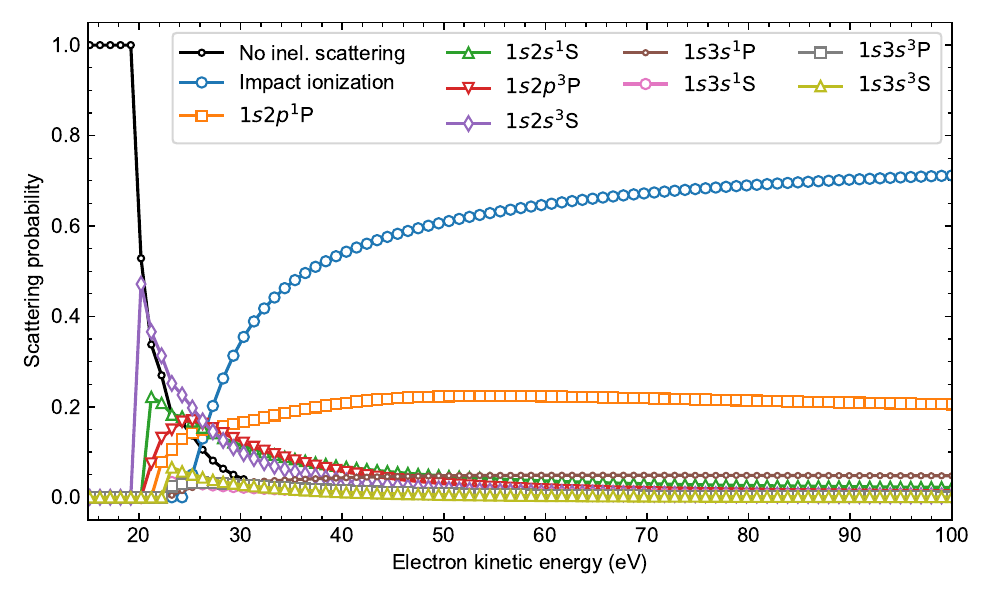}
    \caption{Probabilities for inelastic electron-He scattering into individual channels for electrons traveling through $75$~nm of He derived from tabulated cross sections~\cite{ralchenko2008electron}.
}
    \label{fig:xsecs}
\end{figure}

\subsection{Electron-He inelastic scattering cross sections}

The central component of the simple Monte Carlo simulation introduced in the main text is the energy-dependent inelastic scattering cross sections taken from the literature~\cite{ralchenko2008electron}. The corresponding scattering probabilities for the most important channels are shown in Fig.~\ref{fig:xsecs}. Obviously, below the lowest He excitation at $E_e=19.8$~eV, inelastic scattering is impossible. Just above this threshold, impact excitation of the state 1s2s$\,^3$S is most probable, while at $E_e>27$~eV, impact ionization becomes the most probable channel and the state 1s2p$\,^1$P is the most probable impact excitation process. 

\begin{figure}
    \centering
    \includegraphics{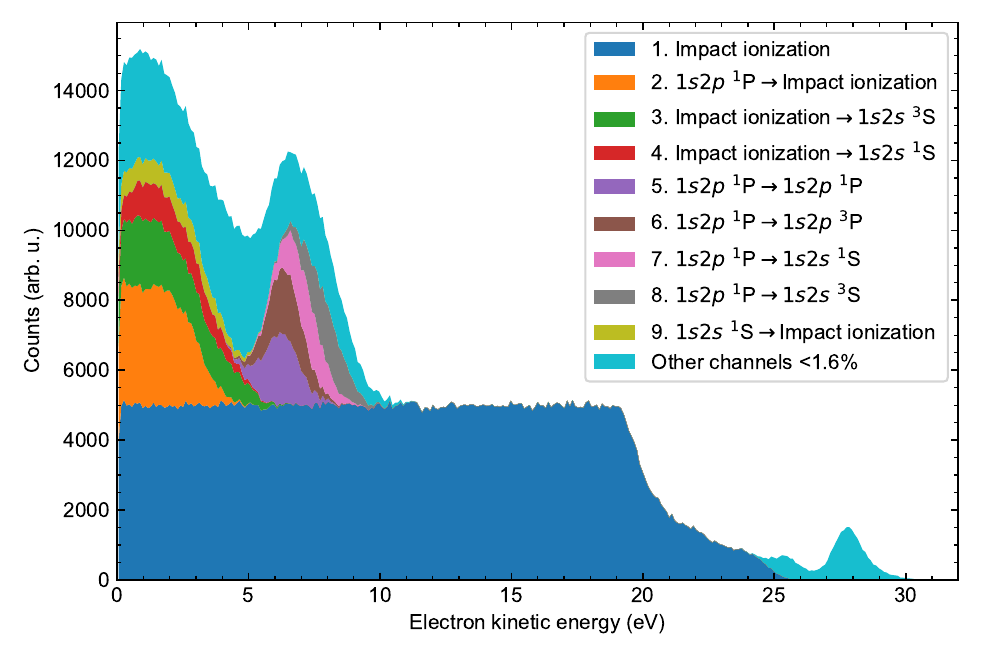}
    \caption{Individual components contributing to the total simulated electron spectrum shown in Fig.~4~b) of the main text as a stacked histogram. The photon energy was $74$~eV, and the He nanodroplet radius was $75$~nm. Legend entries are ordered according to the integrated signal of each component. For multiple inelastic scattering events, the order of the individual scattering channels is indicated by the label.}
    \label{fig:contributions}
\end{figure}

The simulated electron spectra are obtained by binning the electron trajectories according to final electron kinetic energy for selected individual scattering channels; See the resulting histograms in Fig.~\ref{fig:contributions}. Clearly, the edge feature near $E_e=20$~eV arises from impact ionization, whereas the broad peak around $E_e=6$~eV results from two successive collisions where at least one involves the 1s2p\,$^1$P state. All events in which two successive collisions occur contribute to the broad low-energy component at $E_e\lesssim 5$~eV. 

We acknowledge that the simplified assumption of primary and secondary electrons being emitted in the droplet center effectively gives rise to a slightly longer propagation distance inside the nanodroplet than if the initial position of the electron was chosen randomly at any position within the droplet. However, in such a more realistic simulation, random changes of the propagation direction of the electron resulting from both elastic and inelastic collisions would have to be taken into account which would in turn lead to a longer propagation distance and would thus partly compensate for the effect of electron emission in the droplet center in the current implementation. Furthermore, in the experiment, He nanodroplets are generated with different sizes according to a broad size distribution, which would have to be averaged in a more realistic model~\cite{toennies_superfluid_2004}. However, for the sake of keeping the simulation computationally inexpensive and simple enough that it allows us to clearly identify the salient experimental findings -- additional peaks and edges in the electron spectra due to electron energy loss resulting from multiple inelastic scattering -- we refrain from a more detailed description. Consequently, we cannot expect perfect quantitative agreement between experiment and simulation in terms of the relative magnitude of the different spectral features for a given mean He nanodroplet size. Nevertheless, this simulation is instrumental in qualitatively interpreting the newly observed spectral features.

\subsection{Poisson process in space}

\vspace{1em}
Assuming a homogeneous density distribution of He atoms with particle density $n_\mathrm{He}$ and a constant electron-He scattering cross section $\sigma$, the mean free path is $\ell = 1/(\sigma n_\mathrm{He})$. The probability that an electron which has not yet scattered will scatter in the infinitesimal distance interval $[x,x+dx]$ is
\[
dP = \sigma n_\mathrm{He}\, dx = \frac{dx}{\ell}.
\]

Let $S(x)$ denote the survival probability, i.~e., the probability that the particle has propagated a distance $x$ without scattering. Then
\[
S(x+dx) = S(x)\left(1-\frac{dx}{\ell}\right).
\]

\begin{figure}
    \centering
    \includegraphics{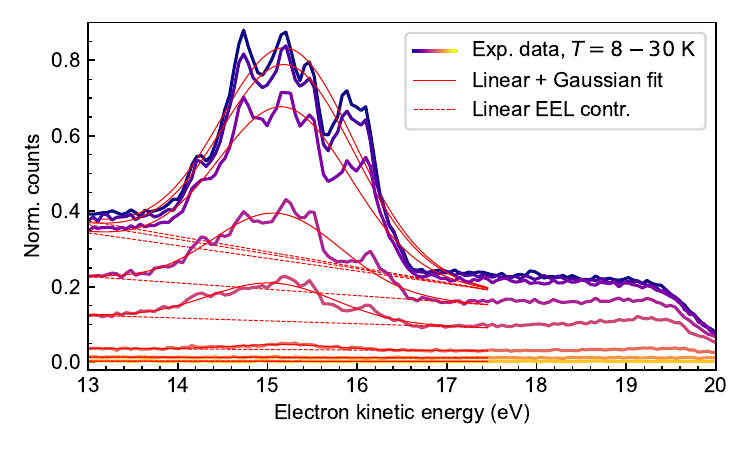}
    \caption{
    Experimental electron spectra in the region of the ICD peak structure (colored thick lines) including fit curves (red lines) that exemplify the fitting procedure. The dashed straight lines depict the contribution from the EEL feature; The solid red lines depict the total fits consisting of the linear component and a Gaussian function to fit the ICD feature. The electron spectra were recorded at $h\nu=72$ eV for various values of the nozzle temperature. }
    \label{fig:icd fit}
\end{figure}

Taking the limit $dx \to 0$ yields
\[
\frac{dS}{dx} = -\frac{S}{\ell},
\]
with solution
\[
S(x) = e^{-x/\ell}.
\]
Thus, the probability of no scattering up to distance $x$ decays exponentially.

The probability density for the first scattering event to occur at position $x$ is
\[
f(x)
=
\frac{1}{\ell} e^{-x/\ell}
=
\sigma n_\mathrm{He}\, e^{-\sigma n_\mathrm{He} x}.
\]
The mean distance to the first scattering event is $\langle x \rangle = \ell$.

\section{Fits of experimental spectra}

This section explains in more detail the fitting procedure of the spectra shown in Fig.~5~a) in the main text to extract the peak integrals shown in Fig.~5~b) and in Fig.~\ref{fig:temps-58-eV}. As the shake-up and ICD peaks overlap with a broad single and multiple EEL peak structure, the individual contributions must be carefully disentangled. The EEL contribution in the ICD energy region $E_e = 14-17$~eV is modeled by a linear function, as a smooth EEL contribution from a single inelastic collision of a photoelectron (due to impact ionization, see Fig.~4 in the main text) is expected. Therefore, the experimental spectra are fitted by the sum of a straight line (EEL) and a Gaussian (ICD), see Fig.~\ref{fig:icd fit}. The dotted red lines depict the linear contributions (EEL) and the solid red lines show the combined fits. 

\begin{figure}
    \centering
    \includegraphics{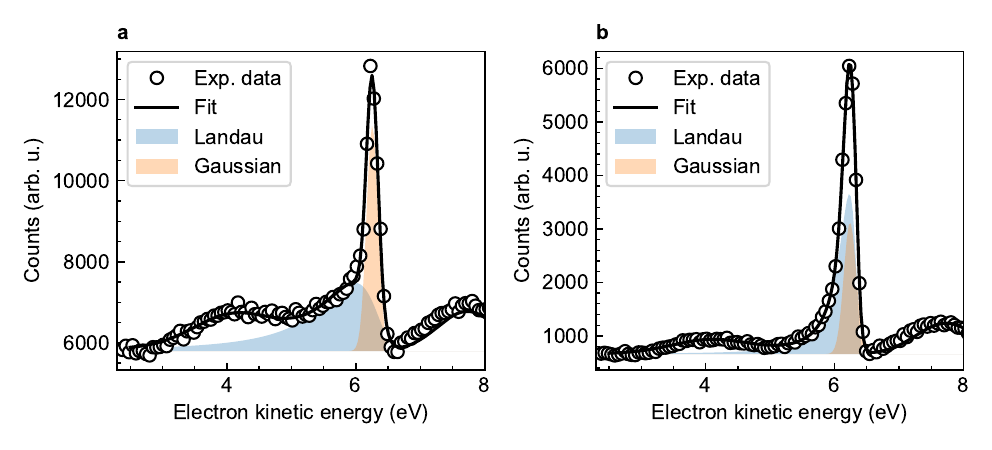}
    \caption{Experimental electron spectra in the region of the satellite peak (circles) including fit curves (solid black lines) that exemplify the fitting procedure. The shaded colored areas depict individual relevant components of the fit. The electron spectra were recorded at $h\nu = 72$ eV for nozzle temperatures $T_0=8$~K (a) and $T_0=14$~K (b).
    }
    \label{fig:shake up fit}
\end{figure}

In the region $E_e < 6.5$~eV, the EEL contribution is modeled by two Gaussians, one to the left and one to the right of the satellite peak due to the shake-up process~\cite{Shcherbinin2017}. The shake-up peak near $E_e = 6$~eV is modeled by the sum of a Gaussian function and a Landau distribution (colored shaded areas in Fig.~\ref{fig:shake up fit}),
\begin{equation}
    \label{eq:Landau}
    f(x) = \frac{1}{\pi} \int_{0}^{\infty}{\exp(-t\log t - xt) \sin(\pi t)}~dt,
\end{equation}
to account for the tail extending toward lower kinetic energy which is more pronounced for lower nozzle temperatures. This tail is likely due to multiple tiny energy losses resulting from elastic electron-He collisions~\cite{asmussen2023electron}.

\section{H\MakeLowercase{e} droplet size dependence in the regime of single collisions}

\begin{figure}
    \centering
    \includegraphics{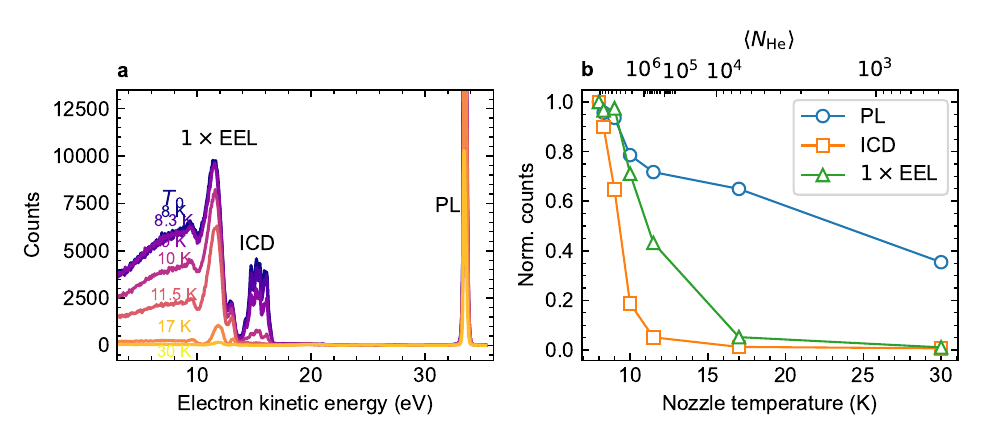}
    \caption{a) Electron spectra recorded at $h\nu = 58$~eV for various nozzle temperatures ranging from $30$~K down to $8$~K. b) Integrals of the most prominent spectral features shown in a), normalized to their respective maximum values, as a function of nozzle temperature, which controls the He nanodroplet size.
    }
    \label{fig:temps-58-eV}
\end{figure}

Fig.~\ref{fig:temps-58-eV} shows the He droplet size dependence of electron spectra recorded at photon energy $h\nu = 58$~eV, i.\,e. below the threshold for two inelastic electron-He collisions, to be compared to Fig.~5 in the main text. Here, the spectra contain, in addition to the photoline (direct photoionization), only peaks resulting from single electron-He collisions (1$\times$EEL) and the characteristic ICD feature at low nozzle temperatures (large He droplets). The peak integrals shown in Fig.~\ref{fig:temps-58-eV}~b) display a similar droplet-size dependence as at $h\nu = 72$~eV (Fig.~5 in the main text). In particular, the EEL feature drops in amplitude faster than the photoline which is indicative of the total density of He atoms in interaction volume; when the mean free path of electrons due to inelastic scattering exceeds the size of the He nanodroplets, the EEL signal vanishes. ICD, relying on multiple inelastic scattering or electron-ion recombination, occurs only at low nozzle temperature where the mean He droplet size exceeds $\sim 10^5$ He atoms per droplet.

\bibliography{Bib}